\documentstyle[aps,pre,epsf]{revtex}

\begin{document}

\pagestyle{empty}

\begin{titlepage}

\begin{center}
{\Large\bf STRESS RELAXATION IN DIBLOCK COPOLYMERS}

\end{center}

\vspace{1cm}

\begin{center}

{\large Adolfo M. Nemirovsky $^*$ and Thomas A. Witten
$^{\dagger}$}
\end{center}
\vspace{1cm}

\begin{center}
{\large\em $^*$ Departament d'Estructura i Constituents de
la Materia, \\Universitat de Barcelona, \\Diagonal 647,
E-08028 Barcelona, Spain. }
\end{center}
\vspace{5mm}

\begin{center}
{\large\em $^\dagger$ The James Franck Institute,\\ The University
of Chicago,\\
5640 S. Ellis Av., Chicago, Illinois 60637, USA.}
\end{center}
\vspace{20mm}

\begin{center}
{\Large\bf Abstract}
\end{center}

\noindent
We study stress relaxation in a strongly segregated
lamellar mesophase of diblock copolymers. We
consider the extreme limit in which chains are
highly stretched and with their junction points confined
to narrow interfaces. A lamella
can be divided into ``stress blobs" at some distance $z$ from
the interface, with well defined local modulus $G(z,\omega)$ at
frequency $\omega$. For sliding (compressional) stress the
total modulus is transmitted in series (parallel) across
the layer. We evaluate the local $G(z,t)$ which shows, for
a given height, a very broad spectrum of relaxation
times.
\vfill \noindent
Published in
{\it Complex Fluids}, conference proceedings for conference in Sitges Spain,
1992, edited by Luis Garrido (Springer-Verlag, Heidelberg, 1993) p.
281.

\end{titlepage}

\section{Introduction}

\indent

Block copolymers made with incompatible blocks form a rich
variety of domain structures. At equilibrium, a dense collection of
monodisperse A-B diblock copolymer chains assemble themselves in
minimum free energy configurations. Symmetric diblock copolymers
with
highly repulsive interactions between A-B segments microphase
separate
into a lamellar order with A-B junction points highly confined into
interfaces to decrease the number of contacts between the two
different
blocks. The width of these lamellae is determined by a balance
between
the reduction of unfavorable contacts between A and B blocks and
the
decrease of chain entropy due to the stretching of the block
polymer.\cite {bates90}

The rheological behavior of diblock copolymers in a lamellar
mesophase is of
great importance due to the numerous practical
applications of
these
systems. The understanding of how polymer domains flow and of how
rheological properties differ from those of the much more familiar
homogeneous melt presents an important and challenging theoretical
problem. Exploratory rheological studies of lamellar domains have
revealed strong departures form the behavior of simple
melts.\cite {bates84}
Theories have begun to grapple with these observations.\cite
{witten90}
\cite {kawasaki90} \cite {obukhov91}
The motion of end-confined chains appears to differ strongly from
that of simpler polymers. Thus, the dynamic response to deformation
in block copolymer domains shows distinctive retardation and
broadening;\cite {bates84} correspondingly, the
disentanglement
dynamics of
end-confined chains are expected to be much different from that of
unconfined polymer melts and solutions.\cite {witten90}
\cite {halperin88} \cite {witten91}

Two cases for the linear response to a small step strain should
be considered \cite {witten92}: a compression of the layers and a
shear tending to
slide
the layers past one another. The block chains have their
translational
motion highly restricted as one chain end is strongly confined to
the
interface, so usual reptation, \cite {doi86} which is believed to
be
the dominant
relaxation mechanism of homogeneous system consisting of long
flexible
polymers chains such as melts (and also, for example, the
disordered
phase of diblock copolymers) is highly suppressed. The
disentanglent
motion of the block chains with one end anchored at the interface
resembles that of star polymers,\cite {doi86} for which relaxation
proceeds by
arm
retraction along the tube (contour length fluctuations).  In
addition, the
block chains are highly stretched along the direction perpendicular
to the lamella, while still retaining their unperturbed (Gaussian)
size in
the directions parallel to the interface.  In fact, relaxation
proceeds at
different rate in different places (depending how far the regions
of interest
are from
the
interface) and, for example, one can define a local
stress modulus $G(z,t)$ with $z$ the distance from the
interface, with regions at different $z$ involving
quite different time scales.
In this contribution we investigate
these distinctive and novel features of heterogeneous materials.

\section{The Equilibrium Lamella}

\indent

Here we restate several relevant properties of copolymer chains as
given, for example, in Refs. \cite{witten90} and
\cite{witten91}
For simplicity, we consider that all interfaces are flat,
parallel to each other and separated by a distance $2h$. We suppose
that
each
surface contains $\sigma$ chains per unit of area. In the
incompressible melt
state each chain fills a volume $V$ proportional to the chain
molecular
weight (and to its chemical length). Hence, the thickness of the
layer $h$
is just given by $h=V\sigma$. Polymers in the melt satisfy
Gaussian
statistics. Thus, the end-to-end distance, which is the
characteristic
dimension of an unperturbed polymer coil, is simply given by
$ R^{2} = V/b$,
with $b$ a microscopic length which only depends on the local
structure of the chain and liquid,  usually of about a few
Angstroms. We
only study here the long chain limit of $V>>b^{3}$.

The block chains ending at the interfaces fill uniformly the
regions between planes so each chain must extend over distances of
the
order of $h$. Stretching the chain a distance $h$
reduces the entropy
of the
random walk thus requiring work of the order of $(h/R)^{2}\;
k_B T$.
We are
interested in high energy interfaces with $(h/R)>>1$.
If a step strain is
applied to the liquid, at very short times it behaves like a rubber
with modulus $G_0$. From
the elastic energy stored by the liquid one can obtain the typical
volume
per entanglement $R_e^{3}$ via $G_0=k_BT/R_e^{3}$.
This $R_e$ is
{\em independent of the molecular weight of
the chain};
hence long chains ($R >> R_e$) are highly entangled.

Since the copolymers are strongly stretched in the lamellar
mesophase, configurational integrals are dominated by the classical
extremun of the action (Hamiltonian),\cite {semenov85}
\cite {milner88}
which corresponds to the most
probable (``classical") configuration of the block extending from
the
interface into a microdomain. In this  approximation, the free
energy
functional for the copolymer blocks in a mesophase domain consist
of
non-interacting Gaussian chains in a self-consistent external
potential (pressure field)
$p(z)$, where $z$ is the distance perpendicular to the interface.
\cite {milner88} \cite {witten90}
This
$p(z)$ is
the free energy cost of bringing a segment to the height $z$ from
some
initial reference position, and it forces the chains to assume a
stretched
state.
We call $v$ the volume
displaced
by the piece of the chain extending from the free end at $z_0$
to the height
$z$. The ``classical" configuration of a chain is given by the
trajectory $z(v)$
of a fictitious Newtonian particle of unit mass in an external
potential
$p(z)$ which starting at some point $z_0$ at ``time" $v=0$
falls to the $z=0$
surface
(interface) in a ``time" $V$.
The ``equal time" property for monodisperse systems gives a
parabolic potential
$p(z) = p_0 (1-(h^{2}/z^{2}))$
with $p_0/k_BT = \pi^{2} h^{2} b/8 V^{2}$
and the chain ``equation of motion"

\begin{equation}
z = z_0 \cos\left(\frac{\pi v}{2V}\right) .
\end{equation}

It remains to specify the
distribution of end positions $z_0$. This distribution is found by
imposing
the constitutive properties of the melt or solution under
consideration.
For example, for a marginal solvent the local volume fraction is
proportional to the pressure $p$.
Instead, for a melt the local volume fraction is
unity
everywhere and we obtain \cite {marko92}

\begin{equation}
\epsilon (z_0) =\frac{\sigma z_0}{h^{2}[1-(z_0^{2}/h^{2})]^{1/2}}
\; ,
\end{equation}

\noindent
where $\epsilon (z_0)$ is the distribution of free ends per unit
area per unit
height.
Chains from opposite interfaces can penetrate each other as long as
work
done against the potential $p(z)$ is of the order of $k_BT$.
Although the energy $p_0 V$ of these chains is typically much
greater than $k_BT$ there is a small region near
the midplane where $p<<p_0$.
In a region of size
$\xi = R (R/h)^{-1/3}$
near this midplane chain segments of size
$\xi$ have energy of order $k_BT$ or less.
In this ``interpenetration
zone" the effects of stretching are unimportant
and the chains from the two opposite
interfaces interpenetrate freely.
In the large $V$ limit of interest,
we have
$h>>R>>\xi>>R_e$,
so segments that live in the interpenetration zone are
still highly entangled.
Figure 1 illustrates the relevant lengths of the
equilibrium lamellar phase of strongly
segregated diblock copolymers.
\begin{figure}[t] 
\epsfxsize=\hsize \advance\epsfxsize by -3in
$$\epsfbox{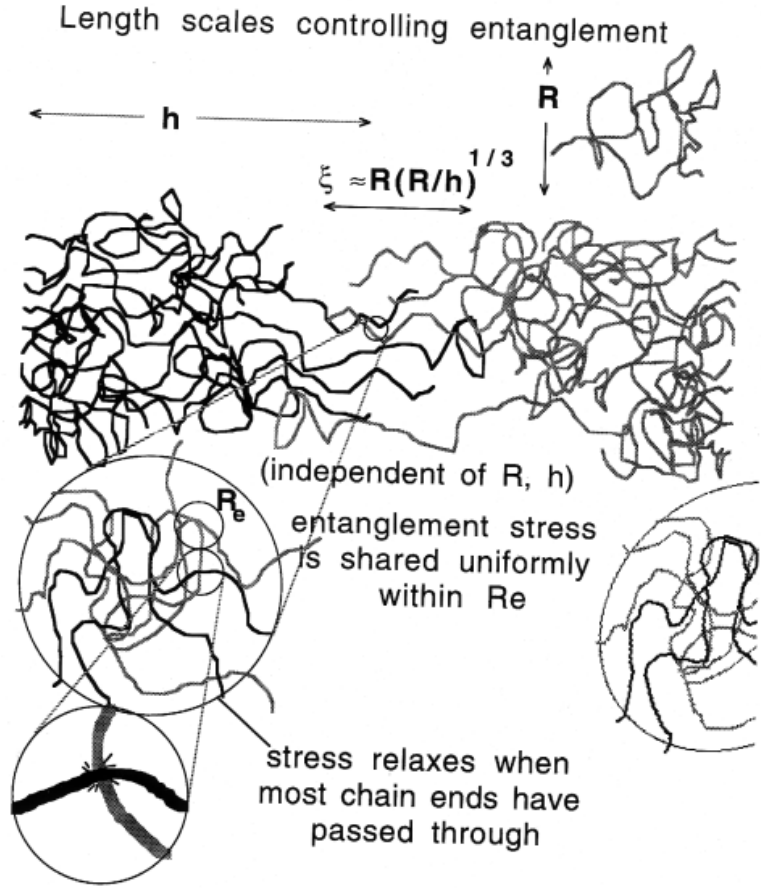}$$
\caption{Relevant lengths in a lamella.}
\label{Fig.1}
\end{figure}

\section{Disentanglement Process in Lamellar
   Diblock
Copolymers}
\indent

After a step-strain is applied to the liquid, the initial shear
distorts the entangled copolymer chains. Since the number of
entanglements per unit of volume is a local property of the melt,
the
initial value of the associated modulus $G_0$ is the same as
that
of a homopolymer melt. Over time the chains disentangle and the
associated distortion dissapears. For the relaxation of
sliding
stress,
the relevant entanglements are those of the interpenetration
region. The
stress is initially held by the entanglements in the
interpenetration zone
as chains from one interface exert force on chains on the other.
Since
there is only weak interpenetration, one would expect that the two
opposite sides disengage relatively fast. On the other hand, over
scales
larger than $R_e$ the stress redistributes more or less uniformly
propagating down to the interfaces, so not only the
interpenetration
region is of relevance in the relaxation mechanism of the sliding
stress.
Compressional stress directly involves the whole chain and not only
interpenetration segments. These entanglements are typically
between
adjacent chains of the same layer, so disentanglement of opposing
layers
is not sufficient to relax the stress.
Thus, compressional relaxation is expected to
procede
slower than sliding stress relaxation.

At the most local level, stress is transmitted along the chains,
but in a larger scale, within a radius several times $R_e$, chains
``collide"
with many other chains redistributing the stress more of less
uniformly
among adjacent chains. At this distance the stress can be treated
as a
smoothly varying quantity. We then divide the liquid into into
imaginary
cubes (``stress blobs") each with a well defined dynamic modulus
$G(z,\omega)$
at a height $z$ and frequency $\omega$.
To evaluate the local modulus $G(z,\omega)$
we
imagine subjecting the single stress blob to a step-strain
experiment illustrated in Figure 2.
The stress relaxes as the initial entanglements disappear. Notice
that
while the stress is defined locally, the disentanglement motion is
not.
To release an entanglement between two chains, it is required that
the
end of either chain pass through the entanglement point. Of course,
the
ends are typically many stress blobs away. Knowing the chains
trailing
volumes $v$ from the height $z$ to their free
ends provides one of the
necesary ingredients to predict how quickly a given stress blob
relaxes.
\begin{figure}[t] 
\epsfxsize=\hsize \advance\epsfxsize by -3in
$$\epsfbox{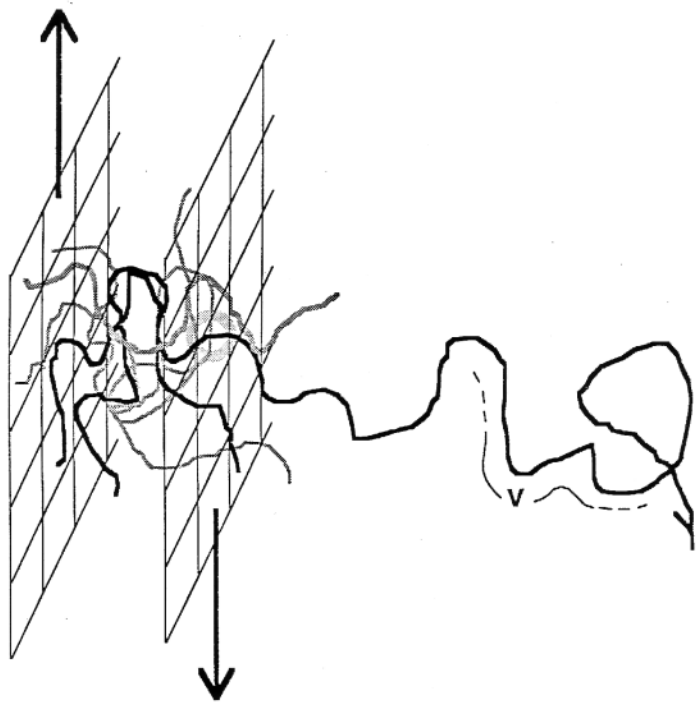}$$
\caption{A blob of size of a few times $R_e$ is subject
to a stress-strain experiment. Although the
stress is defined locally, the free ends of the entangled
chains could be far away.}
\label{Fig.2}
\end{figure}

As  ends of the block polymers are attached to the interface,
reptation
of the chains is highly restricted, and relaxation of stress points
is
expected to proceed by retraction of the tube in analogy with star
polymers. The time distribution for the retraction of segments of
volume $v$ provides a second important ingredient to evaluate the
local
modulus $G(z,t)$. Given the relaxation behavior of these stress
blobs,  one
can readily find how the overall stress relaxes. Layers must be
treated
as a composite material of many sublayers, each with a dynamic
modulus $G(z,\omega)$.
For the sliding shear, the stress is transmitted
in
series across the layer.
During the sliding stress relaxation,
the stresses are constant for all $z$
while the strain is
the average of the strains over
all $z$. The weakest element dominates (blobs
closer to the height h).
In the
compressional
stress, the system is qualitatively like a stack of rubber sheets
which
are clamped together and stretched:
the stress is transmitted in parallel along each
sheet.
The
strongest
moduli dominates and the relaxation time is controlled by the blobs
closest to the interface. Clearly the relaxation is strongly
anisotropic.

The local modulus $G(z,t)$ is then given by

\begin{equation}
G(z,t) = \frac{kT}{R_e^{3}} \int_{0}^{V} dv_1
\int_{0}^{V} dv_2 \; \rho_z(v_1) \rho_z(v_2)
P_{v_1}(t) P_{v_2}(t) \;  .
\end{equation}

The quantity $\rho _z (v)$ is the (normalized) density of
probability that
a
monomer in the stress blob at a height $z$ belongs to a chain
that,
from that height to the free chain end, displaces a volume $v$.
$P_v (t)$ gives
the probability  that a trailing segment of volume $v$ has not yet
retraced
its length in a time $t$.
Then, the integral in Eq. (3) gives the
survival
probability at time $t$ of a stress point at $z$,
assuming that
entanglements are two body processes in accord with the double
reptation model \cite {descloizeaux92}.
In our simplest picture we also assume intra-
and
inter-blob statistical independence. That is, the release of a
stress
point in a blob at height $z$ is independent of similar event at
$z'$,
and that
releasing a constraint within a stress blob will not trigger the
release
of many others.
We also assume that entanglement points are uniformly
distributed (as monomers in the melt are).

\subsection{Height Dependent Trailing Volume Distribution}
\indent

The distribution $\rho _z (v)$
is obtained within the above sketched
``classical
approximation" of the self-consistent field method
appropriate
for
weak excluded volume and moderately high surface coverage for
strongly stretched chains.
The local melt volume fraction
$\phi_z (z_0)$ at a height $z$
contributed by chains
with free ends in the interval $\Delta z_0 = [z_0, z_0 + dz_0]$
(with $z \leq z_0 \leq h$)
is obtained as

\begin{equation}
\phi_z (z_0) dz_0 = \frac{\epsilon(z_0)}{\mid dz/dv \mid}
\; dz_0 \; ,
\end{equation}
\vspace{3mm}
\noindent
where $\epsilon(z_0) dz_0$ given by (2)
is the number of fictitious
particles per unit area falling to
the interface at $z=0$ from the height interval
$\Delta z_0$ and
$\mid dz/dv \mid$ is the particles velocity
at $z$.

\indent
For a given height $z$, when the free end is
at $z_0$, the volume $v$ of the chain trailing segment
is uniquely determined by the equation of motion, Eq (1),
which relates $z$, $z_0$ and $v$. Then, the density of
trailing volume $v$ at $z$ is simply obtained
from (4) changing variables from $z_0$ to $v$
and keeping $z$ fixed. That is, we should replace
$z_0$ by $z_0 (z,v)$ and multiply (4) by the Jacobian
$dz_0 = \mid dz_0/dv \mid dv$.
Then, the
density distribution of
trailing volume $v$ at a height $z$ is given by

\begin{equation}
\rho_z(v) dv = \frac{\epsilon(z_0(z,v))}{\mid
dz/dv \mid }
\mid dz_0/dv \mid  dv\; .
\end{equation}

\vspace{3mm}

\noindent
After performing the algebra and using equations (1), (2) and  (5)
we
obtain

\begin{equation}
\rho_z(v) = \frac{(z/h)}{V \cos(\pi v/2V) \left [\cos^{2}(\pi v/2
V)
-(z/h)^{2} \right]^{1/2}}\; , 0\leq v < v_{max}(z) ,
\end{equation}
\vspace{3mm}

\noindent
with $\rho_z(v)$ vanishing for $v_{max}(z) < v\leq V $.
The function
$\rho _z (v)$ posseses a integrable singularity at
$v_{max} / V= (2/\pi) arcos(z/h)$
indicating that most chains in the stress blob at height
$z$ have a
trailing
segments of volume $v_{max} (z)$.
Clearly, near the middle of the
lamella,
when $z\approx h$, the trailing segments are very short.
Conversely, near
the
interface with $z\simeq 0$, the trailing segments are of the size
of
the
full
chain. Nevertheless, and in contrast with the Alexander
\cite {alexander77}-de Gennes
\cite {degennes76} picture,
there is a distribution of trailing volumes  which, in turn,
gives
rise to a distribution of relaxation times (or frequencies) over an
extended time region. These results which arise from the ``classical
mechanical" description of the stretched polymer chain are slightly
altered when ``quantum" fluctuations are taken into consideration.
(The
major modification are the disappearence of the singularity in
$\rho_z (v)$
and the existence of the penetration zone of width $\xi$).

\subsection{Trailing Volume Survival Probability}

\indent
The calculation of the probability $P_v (t)$ proceeds similarly to
that of Pearson and Helfand \cite {pearson84}
in their study of relaxation of star
polymers. In the tube model, the primitive chain consists of
$\overline{L} /a$
segments
with $\overline{L} = V/b R_e$
the length of the primitive
tube and $a = R_e$ its
step length
($R_e$ is the typical distance between
entanglements).

\indent
In the long chain limit under consideration
there are a large number $\alpha$ of entanglements
per chain, that is $\alpha \equiv R^{2}/R_e^{2} \gg 1$.
We define a variable $x = \overline{L} - L$ which measures
how much the primitive chain has contracted.
$P_v(t)$ is the probability that in time t the primitive tube
of
initial length $\overline{L} \underline{+} \Delta \overline{L}$,
with $\Delta \overline{L}$ the root-mean-square
fluctuation of the tube length, has not yet contracted an
amount
$y = (v/V)\overline{L}$
by an spontaneous thermal fluctuation.
$\Delta \overline{L} =(\overline{L} / (3 \alpha )^{1/2})
\approx (t_R D)^{1/2}$ where $D = k_BT/\zeta$,
$\zeta$ and $t_R$ are, respectively, the diffusion
constant, the chain friction constant, and
the Rouse time.

\indent
We call $p(x,t;x_0;y)$ the density of
probability that the end of the chain is at position $x$
at time $t$, given that initially $(t=0)$ was at
$\mid x_0 \mid \leq \Delta \overline{L}$
and provided that the length of the
primitive tube
was never smaller than $\overline{L} - y$.
It can be obtained by solving a first-passage
problem \cite{pearson84}.
This
probability density is the same as that of a diffusing particle
(the free chain end) in
the
harmonic potential
$U = (3\alpha k_BT/ 2 \overline{L}^{2}) x^{2}$
with an absorbing barrier at y.
Then,
the survival probability at time
$t$ of a tail of volume $v$ is given by
$P_v(t) = \int_{-\infty}^{y} dx \; p(x,t;x_0;y)$,
since, as long as $x \leq y$, the primitive chain has
never contracted by an amount larger
than $y$.
In the limiting case of
$t \gg t_R$ and of $x \gg \Delta \overline{L}$
the probability $P_v(t)$ is obtained as
\cite {pearson84}
\begin{equation}
P_v(t) = \exp (-t/t_v) \; ,
\end{equation}

\noindent
where the time $t_v$ is the typical time at which
a segment of trailing volume $v$ (measured from the
free chain end) has retracted the
tube, and it is given by

\begin{equation}
t_v = t_V \frac {\exp [-(3 \alpha/2) (1 - (v/V)^{2})]}{v/V} \; .
\end{equation}

\noindent
The time $t_V$ at which the whole chain has
retraced the tube is

\begin{equation}
t_V = \frac {\zeta \overline {L}^{2}}{3 \alpha k_B T}
(2\pi/3\alpha)^{1/2} \exp (3\alpha/2),
\end{equation}
\noindent
hence, it is exponentially large in the chain molecular
weight $M$ ($\alpha = R^{2}/R_e^{2} = M / M_e$,
with $M_e$ the entanglement molecular weight).
Clearly, for long chains $t_V$ is much larger than
both,
the Rouse time $t_R \propto M^{2}$,
and
the entanglement time $t_e\propto M^{3.4}$ of linear melts
and concentrated
solutions.
Similar conclusions apply to $t_v$ if $\alpha \gg 1$ and
$v > V /\alpha ^{1/2}$.
Eq. (8) indicates that $t_{v+\Delta v}$ is
{\em several times
larger that} $t_v$
if $\Delta v \simeq 1/\alpha$.
This very strong dependence of $t_v$ on
the volume $v$ of the trailing segment
suggest that the stress relaxation of
the lamellar system is very sensitive
to the trailing volume distribution.
Moreover, from the above discussion
one can conclude that,
for large values of $\alpha$,
the function $P_v(t)$ is
very well approximated by the theta
function

\begin{equation}
P_v (t) = \Theta (v - v(t))\; , \;
\mbox {when} \; \alpha\gg 1 ,
\end{equation}

\noindent
with $v(t)$
obtained by inverting $t_v$ of (8).
This is
because, from Eq.(8),
a given time $t'$ has
associated to it a volume $v'$ such that
$t_{v'} \equiv t'$. Then, if
$v>v'\; (v<v') $, we have $t_v >> t_{v'}\; (t_v << t_{v'})$.

Now  we are ready to evaluate the local
modulus $G(z,t)$ defined in
(3).
Using Eqs. (6) and (10) we obtain

\begin{equation}
G(z,t) = \frac {k_B T}{R_e^{3}} \left [\int_{0}^{V} dv \rho_z(v)
P_v(t) \right ] ^{2}
=\frac {k_B T}{R_e^{3}} \left [\int_{v(t)}^{v_{max}(z)} dv
\rho_z(v) \right ]^{2} ,
\end{equation}

\noindent
which is valid for  $t_R << t < t_V$ . This integral can be solved
in
closed
form as

\begin{equation}
G(z,t) = \frac {k_B T}{R_e^{3}}  \left [1-\frac{2}{\pi} \arctan
\left ( \frac
{(z/h) \sin (\pi v(t)/2V)}{(\cos^{2} (\pi v(t)/2V) -
(z/h)^{2})^{1/2}} \right ) \right ]^{2}\; .
\end{equation}

\vspace{5mm}
\noindent
The logarithm of the local modulus
$\log G(z,t)$ as
a function of $\log(t/t_V)$ is depicted in Figure 3
for various values
of
$(z/h)$.
Clearly, the trailing volume distribution leads to a very broad
(unbounded
in the large $\alpha$ limit) spectrum of relaxation times at a
given height
$z$
above the interface.
\begin{figure}[t] 
\epsfxsize=\hsize \advance\epsfxsize by -3in
$$\epsfbox{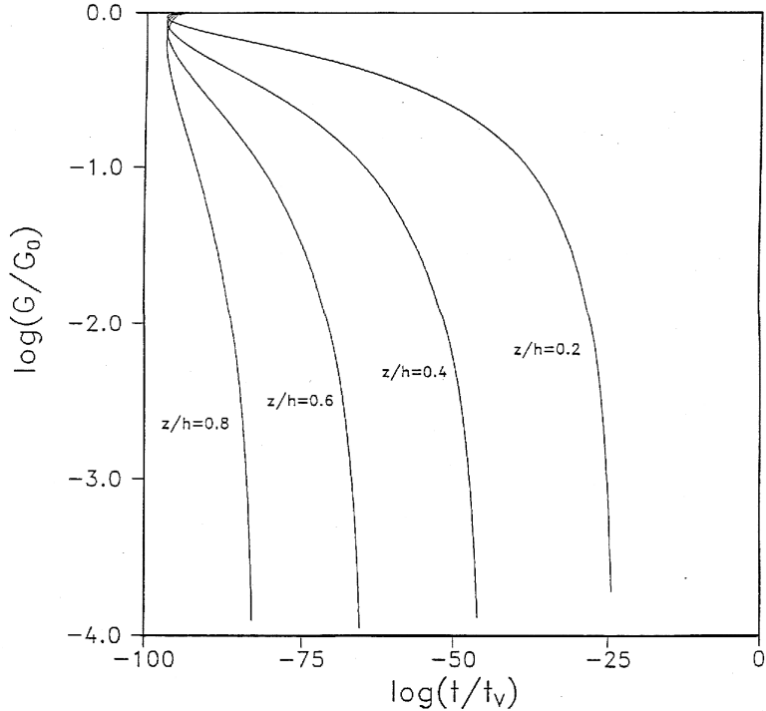}$$
\caption{$\log (G(z,t)/G_0) $ vs. $\log (t/t_V)$ for various values
of $(z/h)$. We have taken $(2 \alpha/3) = 100$.  }
\label{Fig.3}
\end{figure}  

\section{Discussion and Conclusions}
\indent

Stress relaxation in heterogeneous systems
such as the lamellar phase of diblock copolymers
proceeds at different rates in different
spatial regions as shown in Figure 3.
Thus we find that at intermediate times the diblock layer
has three co-existing zones of strongly different rheology:
the region near the midplane is a quickly relaxing liquid;
the regions near each A-B interface are unrelaxed
rubber. Remarkably, this heterogeneity arises from
different parts within a given chain. We would have
obtained similar heterogeneity even if we had neglected
the dispersion of the chains through the layer,
as in Alexander \cite{alexander77}-de Gennes \cite{degennes76}
model. When we include this dispersion, as in Figure 3, we find
in addition a very broad relaxation spectrum,
even at a given height $z$.
This broad spectrum results because of
a) a distribution of tail lengths $v$ at a
given height $z$, and b) the exponential increase
of the relaxation time with the tail length $v$.

Contrary to our expectations,
Eq.(12) predicts that $G(z,t)\equiv 0$ at the midplane $(z = h)$.
This
is because within the ``classical approximation" we have neglected
the penetration zone of width $\xi$. In this region, chains are
Gaussian and hence the distribution of trailing segments is
exponential in the trailing volume $v$; that is
$\rho_{z=0} (v) = \exp (-v/v_\xi)$,
with $v_\xi = b \xi^{2}$ the mean volume of trailing
segments living in the penetration region.
In this region, segments are still
highly entangled ($\xi\gg R_e$)
and stress relaxation proceeds as discussed in
Sec. 3 replacing the trailing volume distribution
(6) by the exponential one. Then, we obtain the
local modulus for a stress blob at the midplane
$G(z=0, t) = (k_BT/R_e^{3}) \exp (-v(t)/v_\xi)$
valid for
$t_R << t < t_{v_\xi}$.

The $G(z,t)$ of (12) is the mean value at
time $t$
of the
stress modulus for a blob at height $z$.
In fact, stress moduli at $(z,t)$ are
distributed according to the binomial
distribution. This is because in a given
blob at a given time, there are many
stress points, each having a probability
$p$ to be alive. The probability $p(z,t)$ is
given by the factor in Eq.(11) that multiplies
the plateau modulus $G_0$.
In principle, the local modulus $G(z,t)$ could
be measured using local probes. On the other
hand, the local modulus is the main ingredient
to evaluate the total modulus for either
sliding or compressional stress.
It is straightforward to find these moduli using
Eq. (12). It is also straightforward to
extend this equation to account for the
interpenetration zone. Our work in this
direction is in progress.

\vspace{7mm}

\begin{tabbing}

\large\bf Acknowledgements
\end{tabbing}

\vspace{3mm}

This work is supported in part by the
Departament d'Ensenyament de la
Generalitat de Catalunya and by
the National
Science Foundation
through Grant No. DMR-88-19860.

\end{document}